\documentstyle[prl,aps,twocolumn,epsfig]{revtex}

\newcommand{\EQ}{\begin{equation}}
\newcommand{\EN}{\end{equation}}
\newcommand{\A}{\begin{array}}
\newcommand{\E}{\end{array}}
\newcommand{\EA}{\begin{eqnarray}}
\newcommand{\EE}{\end{eqnarray}}

\newcommand{\Z}{{\sf Z\!\!Z}}

\newcommand{\simg}{\raisebox{-.6ex}{$\stackrel{>}{\displaystyle{\sim}}$}}

\begin{document}
\twocolumn[\hsize\textwidth\columnwidth\hsize\csname @twocolumnfalse\endcsname
\title{VICINAL SURFACES AND THE CALOGERO-SUTHERLAND MODEL}
\author{Michael L\"{a}ssig}
\address{
Max-Planck-Institut f\"ur Kolloid- und Grenzfl\"achenforschung,
Kantstr.~55, 14513 Teltow, Germany}
\date{\today }
\maketitle

\begin{abstract}

A miscut (vicinal) crystal surface can be regarded as an array of
meandering but non-crossing steps. Interactions between the steps are
shown to induce a {\em faceting transition} of the surface between a
homogeneous Luttinger liquid state and a low-temperature regime
consisting of local step clusters in coexistence with ideal facets.
This morphological transition is governed by a hitherto neglected
critical line of the well-known {\em Calogero-Sutherland model}.  Its
exact solution yields expressions for measurable quantities that
compare favorably with recent experiments on Si surfaces.

\vspace{10 pt}
\vspace{24pt}
\end{abstract}


\vfill
] 
\narrowtext

Miscutting a crystal at a small angle with respect to one of its
symmetry planes produces a {\em vicinal surface} \cite{Wortis.review}.
It often consists of a regular array of terraces separated by
monoatomic steps.  The steps meander by thermal activation but they do
not cross or terminate; their density is determined by the miscut
angle. This picture, the well-known {\em terrace-step-kink
model}~\cite{Jayapakrash.TSK}, neglects the formation  of islands,
voids, and overhangs on the surface, and is hence expected to be valid
below the roughening transition of the ideal facet. In the simplest
approximation, such steps are modeled as the world lines of {\em free
fermions} moving in one spatial dimension $r$ and imaginary time $t$, thus
taking into account the no-crossing constraint through the Pauli
principle \cite{fermions,DenNijs.DG}. While the free fermion model is
sometimes a qualitatively satisfactory approximation \cite{Bartelt.freef},
it has become clear that {\em interactions} between the steps can
induce phase transitions that change the surface morphology
drastically~\cite{Jayapakrash.TSK}. From a theoretical point of view,
models of interacting fermions are important realizations of
two-dimensional euclidean field theories, some of which are exactly
solvable. For example, steps with short-ranged interactions can be
mapped onto the Thirring model or, in the more complex case of
reconstructed surfaces, onto the Hubbard model
\cite{BalentsKardar.Hubbard}.

Interactions between steps are produced by a variety of physical
mechanisms~\cite{RedfieldZangwill}. For example, elastic forces lead to
a long-ranged mutual {\em repulsion} that decays as $r^{-2}$ with the
step separation $r$ \cite{AndreevKosevich}.  Short-ranged interactions
(including all forces that decay faster than $r^{-2}$) can be of either
sign. Using scanning tunneling microscopy on Cu surfaces, Frohn et
al.~\cite{FrohnAl.steps} have found evidence for step-step {\em
attractions} that decay over distances of a few lattice spacings. In a
beautiful series of X-ray scattering experiments, Song and
Mochrie~\cite{SongMochrie} have recently discovered an important
manifestation of attractive forces on miscut Si(113) surfaces in
equilibrium. At sufficiently high temperatures ($ T \simg 1300 {\rm
K}$), a surface of miscut angle $\theta_0$ is a homogeneous ensemble of
fermionic steps whose local density $\rho (r,t)$ has the expectation
value $ \langle \rho (r,t) \rangle = \tan \theta_0 \equiv \rho_0 $ and
somewhat smaller fluctuations $ \langle \rho (r,t) \rho (r',t') \rangle
$ than expected for free fermions. As the temperature is lowered,
however, the fluctuations increase substantially, until a {\em faceting
transition} occurs at a temperature $ T^\star (\rho_0) \approx 1200
{\rm K}$. Below that temperature, the attractive forces cause the steps
to cluster locally.  Hence the surface splits up into domains of an
increased and {\em temperature-dependent}~\cite{tempdep} step density $
\langle \rho(r,t) \rangle = \bar \rho (T) > \rho_0 $ alternating with
step-free (113)-facets ($ \langle \rho (r,t) \rangle = 0 $).  A
critical temperature $T_c = 1223 {\rm K}$ is identified from the
extrapolation $\bar \rho (T_c) = 0$.

This Letter is devoted to a theoretical analysis of stepped surfaces
with both long- and short-ranged interactions. By mapping the step
ensemble onto an exactly solvable fermion model, it is shown that a
long-ranged repulsion and a short-ranged attraction can conspire to
produce a quite complex temperature dependence of the surface
morphology, including a faceting transition in {\em quantitative}
agreement with the experimental findings of ref.~\cite{SongMochrie}. I
obtain three temperature regimes characterized by qualitatively
different step configurations (see Fig.~1):
\begin{figure}[b]
\epsfig{figure=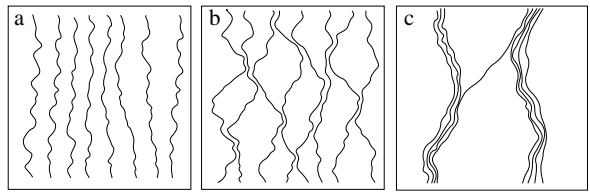,width=1in,angle=270}
\vspace{12pt}
\caption{
Typical configurations of non-crossing (fermionic) steps coupled
by inverse-square and short-ranged forces.
(a)~High-temperature regime  ($ T > \tilde T (\rho_0) $).
(b)~Critical regime ($ \tilde T (\rho_0) > T > T^\star (\rho_0)  $).
(c)~Faceted regime ($T < T^\star (\rho_0)$).
}
\end{figure}
(a)~Well above the critical temperature $T_c$, the steps are dominated
by the no-crossing constraint and the long-ranged repulsion. Hence, they are
well separated from each other with relatively small fluctuations. 
Below a crossover temperature $\tilde T (\rho_0)$, the short-ranged 
attraction becomes important.
(b)~In the critical regime close to $T_c$, the probability of a step
being close to one of its neighbors is substancially enhanced. This
goes along with increased step fluctuations and a broader distribution
of terrace widths.
(c)~Below the faceting temperature $T^\star (\rho_0) < T_c$, the steps
form local bundles of density $ \langle \rho(r,t) \rangle > \rho_0 $.
On average, the distance between two neighboring bundles is larger than
the width of an individual bundle. The fluctuations of these
``composite'' steps are smaller than those of individual steps.

Specifically, I consider a system of $p$ fermionic lines $r_i (t)$ governed
by the effective action
\begin{equation}
S = \frac{1}{T} \int {\rm d}t
	   \left [ \frac{1}{2} \sum_{i=1}^{p} \dot{r}_i^2 +
		   \sum_{i < j} (g \omega_a (r_{ij}) +
		                   h \delta_a (r_{ij}) )
           \right ]  ,
\label{S}
\end{equation}
where $ \dot{r}_i \equiv {\rm d}r_i / {\rm d}t $ and $ r_{ij} \equiv r_i
- r_j$. The action contains kinetic terms with a line tension
normalized to 1, ``contact'' forces $\delta_a (r)$ of
microscopic range $a$~\cite{regularization}, and an ``equal-time''
approximation $\omega_a (r) = r^{-2}$ to  the elastic interactions for
$ |r| > a$ \cite{equaltime}. The universal properties can be expressed
in terms of the rescaled coupling constants $g_0 \equiv g/T^2$ and $h_0
\equiv h/T^2$.  In the limit $\rho_0 a \ll 1$ of small miscut angles,
this system can be mapped onto the {\em Calogero-Sutherland
model}~\cite{CalogeroSutherland}, an exactly solvable continuum theory
well known in the context of the fractional quantum Hall effect and
random matrix theory. Its two branches of solutions are labeled by the
parameter
\begin{equation}
\lambda^\pm (g_0) = \frac{1 \pm \sqrt{ 1 + 4g_0 }}{2} \;.
\label{lambda}
\end{equation}
The strong temperature dependence of the surface morphology described
above is shown to arise from crossover phenomena between these two
branches of solutions. In the high-temperature regime,
the surface is governed by the solution  $\lambda^+ (g/T^2)$. At the
critical temperature $T_c$ (implicitly given by 
$h/ T_c^2 = h_0^- (a, g/T_c^2)$, 
where $h_0^- (a, g_0)$ is a nonuniversal function), the surface
scales according to the solution $\lambda^- (g_0^c)$ (as long as $g_0^c
\equiv g / T_c^2 < 3/4$; beyond that point, this branch of solutions
ceases to exist). The solution $\lambda^- (g_0^c)$ also determines the
singular density dependence of the crossover temperature 
$\tilde T (\rho_0)$ and the faceting temperature $T^\star (\rho_0)$ (i.e.,
of the size of the critical regime).   
Hence, a number of observables at different temperatures are predicted
in terms of the single nonuniversal parameter $g_0^c$, which allows
direct comparison with experiments (see the discussion at the end of
this Letter).

The solutions of the Calogero-Sutherland model labeled by $ 0 < \lambda
< \infty$ are known to describe a line of Luttinger liquid critical
points \cite{KawakamiYang} that contains the self-dual point $\lambda =
2$, the free fermion point $\lambda = 1$ and the Kosterlitz-Thouless
point $\lambda = 1/2$.  Notice, however, that for repulsive long-ranged
forces ($g >0$), the root $\lambda^-(g_0^c)$ is negative. Solutions of
the Calogero-Sutherland model with $\lambda < 0$ have not been
discussed before as they were deemed unphysical. The solutions labeled
by $ 0 > \lambda > -1/2$ form a new line of Luttinger liquid critical
points; faceting on vicinal surfaces seems to be their first
realization.  This line is the analytic continuation of the line  $ 0 <
\lambda < \infty$ beyond the free boson point $\lambda = 0$ (see
Fig.~2), and it terminates at its Kosterlitz-Thouless point $\lambda =
- 1/2$. There is another closely related physical manifestation of the
solutions $\lambda^- (g_0^c)$ in the particular case $p = 2$, where the
action (\ref{S}) is a model for two interfaces in a two-dimensional
system in the so-called intermediate fluctuation regime
\cite{LipowskyNh.wetting}.  In these systems, the well-known line of
{\em wetting} critical
points~\cite{LipowskyNh.wetting,Lipowsky.parabolic} turns out to
correspond to that branch of solutions. The wetting transition is of
second order for $\lambda > -1/2$, but of first order for $\lambda \leq
-1/2$ \cite{LipowskyNh.wetting,Lipowsky.parabolic}.

\begin{figure}
\epsfig{figure=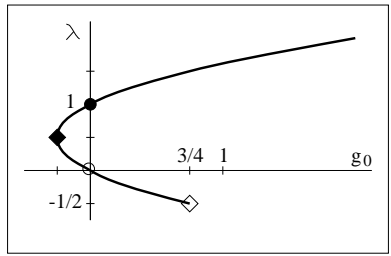,width=1.3in,angle=270}
\vspace{12pt}
\caption{
The two branches $\lambda^+ (g_0)$ and $\lambda^- (g_0)$ of the
Calogero-Sutherland model. On vicinal surfaces, they govern the
high-temperature regime and the faceting transition, respectively.
The solutions $0 < \lambda < \infty$ and $0 > \lambda > - 1/2$
form two distinct lines of Luttinger liquid critical points. Special
points are $\lambda = 1$ (free fermions), $\lambda = 0$ (nonrelativistic
free bosons) and $\lambda = \pm 1/2$ (Kosterlitz-Thouless points).}
\end{figure}

To derive these results, it is convenient to regard the ensemble of steps
as a many-body quantum system. The Hamiltonian of this system,
\begin{equation}
H = - \frac{1}{2} \int {\rm d} r \, 
    \psi^\dagger (r,t) \partial_r^2 \psi (r,t)
    + g_0 \Omega_a (t) + h_0 \Phi_a (t) ,
\label{H}
\end{equation}
acts on $p$-particle states; its form is determined by the action (\ref{S}).
$\psi$ and $\psi^\dagger$ are anticommuting fields;
$ \Omega_a (t) \equiv
  \int {\rm d}r {\rm d}r' \rho (r,t) \omega_a (r - r') \rho (r',t) $ and
$\Phi_a (t) \equiv
  \int {\rm d}r {\rm d}r' \rho (r,t) \delta_a (r - r') \rho (r',t) $ are
the long- and short-ranged interactions written in terms of the density
operator $ \rho (r,t) \equiv \psi^\dagger (r,t) \psi (r,t)$. In a system of
finite width  $L$ with periodic boundary conditions and the periodic potential
$ \omega_a (r) = (\pi^2 / L^2) \sin^{-2} (\pi r / L) $ (for $ a < r < L-a) $, 
the two-particle ground state takes the exact form \cite{CalogeroSutherland}
\begin{equation}
\Psi_2 (r_{12}) = \sin^\lambda (\pi r_{12} / L)
\label{Psi2}
\end{equation}
(for $ a < r_{12} < L - a)$, with $\lambda$ given by Eq.~(\ref{lambda}).
A contact potential of fixed strength $h_0 = h_0^\pm (a, g_0)$
is required to match the wave function (\ref{Psi2}) with the fermionic 
boundary condition $\Psi_2 (0) = \Psi_2 (L) = 0$ \cite{singularity}. 
In the limit $a \to 0$, the $p$-particle ground state is the simple product
\begin{equation}
\Psi_p (r_1, \dots, r_p) = \prod_{i<j} \Psi_2 (r_{ij}) \;,
\label{Psip}
\end{equation}
provided $\lambda > - 1/2$ (otherwise this wave function is not normalizable).
An integrable continuum model emerges, known as the Calogero-Sutherland
model~\cite{CalogeroSutherland}.

Hence, for a given value $-1/4 < g_0 < 3/4$, the Hamiltonian (\ref{H})
defines two different continuum theories, corresponding to the branches
$\lambda^\pm (g_0)$ in Eq.~(\ref{lambda}). The stability of these solutions
in the thermodynamic parameter space $(g_0, h_0)$
can be studied perturbatively, using the methods of
refs.~\cite{Roughening,Fermions,Altenberg} (where the reader is referred
to for more details). 
The expansion of the dimensionless $p$-particle ground state energy
$F_p \equiv L^2 E_p$ about the branch point $\lambda^\pm = 1/2$ has the form
\[
F_p (g'_0, h'_0) - F_p (0,0) = \frac{(-1)^{M + N}}{M! N!}
                               \sum_{M + N \geq 1} F_{M,N} \, {g'_0}^M {h'_0}^N 
\]
with
$g'_0 \equiv g_0 + 1/4$,
$h'_0 \equiv h_0 - h_0^- (a, -1/4)$, and
\[
F_{M,N}  = \rho_0^{-2} 
           \int \prod_{i = 2}^{M + N} {\rm d}t_i \, \langle 
                \prod_{i = 1}^M \Omega_a (t_i)
                \prod_{i = M + 1}^{M + N} \Phi_a (t_i)  \rangle  
\]
otherwise. The brackets $\langle \dots \rangle$ denote connected
expectation values in the unperturbed $p$-particle ground state
(\ref{Psip}) at $\lambda = 1/2$. In the limit $a \to 0$, the
coefficients $F_{M,N}$ develop logarithmic singularities. With an
appropriate normalization of the operators  $\Phi_a$ and $\Omega_a$,
one finds at the lowest orders
\begin{eqnarray*}
F_{1,0} & = &  \rho_0^{-2} \langle \Omega_a  \rangle 
               = s \rho_0^{-2} \langle \Phi_a \rangle 
               + O(s^0) \;, 
\\
F_{0,2} & = &  \rho_0^{-2} 
               \int  {\rm d} t \,
               \langle  \Phi_a (0) \Phi_a (t) \rangle 
               =  2 s \rho_0^{-2} 
               \langle \Phi_a  \rangle + O(s^0) \;, 
\end{eqnarray*}
and hence
\begin{eqnarray*}
F_p (g'_0, h'_0) - F_p (0,0) &  = &  
  - \rho_0^{-2} \langle \Phi_a \rangle  ( h'_0 + s (g'_0 - {h'_0}^2))  
 \\     & &          + O(s^0, {g'_0}^2, g'_0 h'_0, {h'_0}^3) \;,
\end{eqnarray*}
where $ s \equiv - \log (\rho_0 a) $. Up to this order, the
singularities can be absorbed into the renormalized coupling constant
$h_R \equiv h'_0 (1 - s h_R) + s g'_0 + \dots$,
while no renormalization is needed
for $g_0$. This leads to the parabolic flow equations
\begin{equation}
\frac{{\rm d}}{{\rm d} s} \, g_0 = 0 \;, \hspace{1cm}
\frac{{\rm d}}{{\rm d} s} h_R = \frac{1}{4} + g_0 - h_R^2 \;,
\label{beta}
\end{equation}
which are independent of $p$. 
They have first been obtained by functional renormalization group
methods for $p = 2$ \cite{Lipowsky.parabolic}  and have  been derived for
arbitrary $p$ in refs.~\cite{KoStrMukBh.LR,KinzelbachLassig.LR}.  It is
possible to check that $F_{1,0}$ and $F_{0,2}$ contain the only
primitive singularities of the perturbation series. The renormalization
group equations (\ref{beta}) are thus exact to all orders in a minimal
subtraction scheme.

There are two lines of fixed points, $h_R^\pm = \pm \sqrt{1 + 4 g_0}/2$.
The renormalization group eigenvalue of temperature variations 
(or variations of $h_R$), 
\begin{equation}
y^\pm (g_0) = \frac{1}{2} \left. \frac{\partial}{\partial h_R} 
              \right |_{ h_R^\pm (g_0) } 
              \left ( \frac{{\rm d} h_R}{{\rm d} s} \right ) 
            = \mp \frac{\sqrt{1 + 4 g_0}}{2} \; ,
\label{y}
\end{equation}
also governs the scaling of the contact operator $\Phi_a$,
e.g., $ \langle \Phi_a \rangle^\pm (g_0) \sim \rho_0^{ 2 (1 - y^\pm
(g_0)) } $.  This is precisely the scaling of $\langle \Psi_p | \Phi_a
| \Psi_p \rangle $ obtained from the exact solution (\ref{Psip}) with
$\lambda$ given by (\ref{lambda}). The two lines of fixed points
$h_R^\pm (g_0)$ can thus be identified with the two branches of
solutions $\lambda^\pm (g_0)$ of the Calogero-Sutherland model.

At the fixed points $\lambda^- (g_0^c)$, temperature variations are a 
relevant perturbation.
(For $p = 2$, these fixed points govern the wetting transition at the
critical temperature $T_c$ \cite{Lipowsky.parabolic}.)
Above $T_c$, Eq.~(\ref{beta}) yields a crossover to the stable branch
of solutions $\lambda^+ (g_0)$. The crossover temperature is given by
$\tilde T (\rho_0) - T_c \sim \rho_0^{2 y^- (g_0^c)}$. Below $T_c$, 
the renormalized coupling $h_R$ tends to $-\infty$ under the flow
(\ref{beta}), indicating an instability of the step ensemble with
respect to the formation of local bundles. These bundles have a 
characteristic line density 
\begin{equation}
\bar \rho (T) \sim (T_c - T)^{1 / 2 y^- (g_0^c)} \;. 
\label{rhobar}
\end{equation}
The clustering becomes visible if $\bar \rho (T) > \rho_0$, i.e.,
for $T < T^\star (\rho_0)$ with 
$ T_c - T^\star (\rho_0) \sim \rho_0^{2 y^- (g_0^c)}$.
On average, a bundle consists of $n \approx 20$ lines~\cite{SongMochrie,n}
and has a width $n / \bar \rho(T)$. Two neighboring bundles at a typical
distance $n / \rho_0$ have an exponentially small overlap 
$\sim \exp ( - \bar \rho (T) / \rho_0 )$; they can thus be approximated as 
stable composite steps with nearly step-free facets in between. 

From the above discussion, it is clear that the integrability of this
system is tied to scale invariance at distances $a \ll r \ll
\rho_0^{-1}$. Along the crossover between the critical and the
high-temperature regimes, the two-particle wave function does not
have the simple power-law asymptotics $\Psi_2 (r) \sim |r|^\lambda$
as in (\ref{Psi2}), and consequently, the product ansatz (\ref{Psip}) 
breaks down. What happens to scale 
invariance at distances $r \gg \rho_0^{-1}$? It has been shown that 
any solution of the Calogero-Sutherland model with $\lambda > 0$
describes a Luttinger liquid: it belongs to the universality 
class of the Gaussian model with action
\begin{equation}
S_G = \frac{\gamma}{4 \pi} 
      \int ({\bf \nabla} h ({\bf r}) )^2 {\rm d}^2 {\bf r}
\label{SG}
\end{equation}
and stiffness $ \gamma = \lambda / 2$ \cite{KawakamiYang}. Here $ {\bf
r} \equiv (r, v_F t)$, where $v_F = 2 \pi \lambda \rho_0$ is the Fermi
velocity, and $h({\bf r})$ is a coarse-grained surface
height variable.  Using the Bethe ansatz, one finds the low-lying
finite-size excitations $\Delta E_{e,m} = 2 \pi v_F x_{e,m} / L$ in
terms of the scaling dimensions $x_{e,m} = (e^2 \gamma + m^2 / \gamma)
/ 2$ of the Gaussian vertex operators ${\cal O}_{e,m}$ ($e,m \in \Z$)
\cite{KawakamiYang}. It is then easy to show that any solution with
$\lambda < 0$ is also a Luttinger liquid with $\gamma = - \lambda / 2$,
since the transformation $\lambda \to - \lambda$ acts as the symmetry
${\cal O}_{e,m} \to {\cal O}_{e,-m}$ on the Gaussian operator algebra.
Thus conformal field theories with central charge $c = 1$ govern the
steps at $T_c$ {\em and}\ in the high-temperature regime - and thus
along the entire crossover by virtue of the $c$-theorem
\cite{Zamolodchikov.cth}. It follows that the system is a Luttinger
liquid at all temperatures $T > T_c$ \cite{redundant}.  This property
extends to the entire critical regime, where $\gamma$ can be written in
scaling form, $\gamma(T, \rho_0) = \Gamma ( (T - T_c) \rho_0^{ -2 y^-
(g_0^c)} )$.  Below $T^\star (\rho_0)$, one expects an effective action
similar to (\ref{S}) for the composite steps;  the system is then still
a Luttinger liquid.

We now turn to some experimentally measurable consequences of this
theory. \newline
(a) The ground state wave function (\ref{Psip}) yields immediately the
step density correlation function
\begin{equation}
\langle \rho (0,t) \rho (r,t) \rangle \sim \Psi_2^2 (r) 
                                      \sim r^{2 \lambda}
\end{equation}
in the limit $ |r| \ll \rho_0^{-1} $, where multi-particle effects can
be neglected. Two steps at distance $r$ from each other enclose a
terrace of width $r$ if no further steps are in between; this condition
is also negligible for $r \ll \rho_0^{-1}$. Hence the terrace width
distribution, which can be measured by surface scanning techniques, has
the same short-distance tail $ \sim r^{2 \lambda}$
\cite{Bartelt.freef}. In the high-temperature regime ($\lambda =
\lambda^+ (g / T^2) > 1$), short terraces are rare, while at $T_c$
($\lambda = \lambda^- (g_0^c) < 0$), they are abundant. \newline 
(b) The surface stiffness $\gamma$ is measurable as universal pre-factor 
of the height difference correlation function
\begin{equation}
C(r) \equiv \frac{1}{2} \langle (h(0,t) - h(r,t))^2 \rangle 
     \simeq \gamma^{-1} \log |\rho_0 r| + \dots 
\end{equation}
on scales $ |r| \gg \rho_0^{-1}$. 
$\gamma$ shows a characteristic temperature dependence. Its
high-temperature asymptotic value is $\gamma = 1/2$. As the temperature 
is lowered, it first {\em increases} as 
$\gamma = \lambda^+ (g / T^2) / 2 = \lambda^+ (g_0^c T_c^2 / T^2) / 2$,
then {\em decreases} in the critical regime 
(taking the value $\gamma = - \lambda^- (g_0^c) / 2 < 1/4$ at $T_c$), and 
again sharply {\em increases} below $T^\star (\rho_0)$ to values $\gamma > 
1/2$ for composite steps. Most aspects of this pattern have been observed
\cite{SongMochrie}, but the measurements are not yet conclusive in the 
high-temperature regime. In the critical regime, the data have been fit to
a power law $\gamma \sim (T - T_s (\rho_0))^{- \kappa}$ with 
$T_s (\rho_0) < T^\star (\rho_0)$ \cite{SongMochrie}, which corresponds 
to a singularity in the scaling function $\Gamma$.  \newline
(c) The temperature dependence of the cluster density (\ref{rhobar}) agrees 
with the measurement 
$\bar \rho (T) \sim (T_c - T)^{0.42 \pm 0.10}$ \cite{SongMochrie}
if $g_0^c \approx 3/4$. \newline
(d) The temperature differences $\tilde T (\rho_0) - T_c$, 
$T_c - T^\star (\rho_0)$, and $T_c - T_s (\rho_0)$ depend on the step 
density as $\rho_0^{2 y^- (g_0^c)}$. This is also consistent with the 
data for $g_0^c \approx 3/4$.

In summary, the Calogero-Sutherland model has been applied to
interacting steps on vicinal surfaces. The thermodynamic complexity 
of this system arises from the interplay of the two branches of
integrable solutions. It will be of interest whether this mechanism 
also plays a r\^ole in other realizations of the Calogero-Sutherland
model.
  
I thank J.L. Cardy, V. Korepin, R. Lipowsky, S.M.~Mochrie, and M.~Zirnbauer
for useful discussions. In particular, I am grateful to
S.M.~Bhattacharjee for his contribution at the initial stage of this
work.


\begin{references}



\bibitem{Wortis.review}
For a review, see M. Wortis, in {\em Chemistry and Physics of Solid
Surfaces}, Vol.~7, Springer-Verlag, Berlin, 1988.



\bibitem{Jayapakrash.TSK}
C. Jayaprakash, C. Rottmann, and W.F. Saam, Phys. Rev. B 30 (1984), 6549.



\bibitem{fermions}
P.G. de Gennes, J. Chem. Phys. 48 (1968), 2257;
V.L. Pokrovski and A.L. Talapov, Phys. Rev. Lett. 42 (1979), 65.



\bibitem{DenNijs.DG}
For a review of fermionic methods, see
M. den Nijs, in {\em Phase Transitions and Critical Phenomena}, Vol.~12,
ed. by C. Domb and J.L. Lebowitz, Academic, London, 1989.



\bibitem{Bartelt.freef}
N.C. Bartelt, T.L. Einstein, and E.D. Williams, Surf. Sci. Lett. 240
(1990), L591;
B. Jo\'os, T.L. Einstein, and N.C. Bartelt, Phys. Rev. B 43 (1991), 8153.



\bibitem{BalentsKardar.Hubbard}
L. Balents and M. Kardar, Phys. Rev. B 46 (1992), 16031.



\bibitem{RedfieldZangwill}
A.C. Redfield and A. Zangwill, Phys. Rev. B 46 (1992), 4289.



\bibitem{AndreevKosevich}
A.F. Andreev and Y.A. Kosevich, Zh. Eksp. Teor. Fiz. 81 (1981), 1435
[Sov. Phys. JETP 54 (1881), 761].



\bibitem{FrohnAl.steps}
J. Frohn et. al., Phys. Rev. Lett. 67 (1991), 3543.



\bibitem{SongMochrie}
S. Song and S.G.J. Mochrie, Phys. Rev. Lett. 73 (1994), 995;
Phys. Rev. B 51 (1995), 10068.



\bibitem{tempdep}
Below 1134K, the step clusters turn into (114)-facets~\cite{SongMochrie}, 
i.e., $\bar \rho$ becomes independent of temperature. This is a more 
familiar kind of faceting. 



\bibitem{regularization}
The regularization $\delta_a (r)$ of the contact potential is familiar
in field theory as point splitting. It is necessary since a naive
contact term $\delta (r)$ vanishes identically due to the Pauli principle.



\bibitem{equaltime}
This approximation is justified a posteriori by the fact that it
preserves the rotational (and even the conformal) symmetry of the
theory at large distances. Indeed, it is the leading term in a gradient
expansion, all subsequent terms of which are irrelevant under
renormalization.



\bibitem{CalogeroSutherland}
F. Calogero, J. Math. Phys 10 (1969), 2191 and 2197;
B. Sutherland, J. Math. Phys. 12 (1971), 246 and 251;
Phys. Rev. A 4 (1971), 2019.



\bibitem{KawakamiYang}
N. Kawakami and S.-K. Yang, Phys. Rev. Lett. 67 (1991), 2493.



\bibitem{LipowskyNh.wetting}
R. Lipowsky and T.M. Nieuwenhuizen, J. Phys. A 21 (1988), L89.



\bibitem{Lipowsky.parabolic}
R. Lipowsky, Phys. Rev. Lett. 62 (1989), 704;
Physica Scripta T 29 (1989), 259.



\bibitem{singularity}
As $a \to 0$, this strength diverges, $ h_0^\pm (a, g_0) \sim a^{-1}$.



\bibitem{Roughening}
M. L\"assig and R. Lipowsky, Phys. Rev. Lett. 70 (1993), 1131.



\bibitem{Fermions}
M. L\"assig, Phys. Rev. Lett. 73 (1994), 561.



\bibitem{Altenberg}
M. L\"assig and R. Lipowsky, in {\em Fundamental Problems of
Statistical Mechanics VIII}, Elsevier, Amsterdam, 1994. 



\bibitem{KoStrMukBh.LR}
S. Mukherji and S.M. Bhattacharjee, unpublished;
E.B. Kolomeisky and J.P. Straley, Phys. Rev. B 46 (1992), 13942;
Phys. Rev. Lett. 73 (1994), 1648.



\bibitem{KinzelbachLassig.LR}
H. Kinzelbach and M. L\"assig, unpublished (see ref.~\cite{Altenberg}).



\bibitem{n} 
To determine $n$ from the theory, further nonlocal step interactions
have to be taken into account, which is beyond the scope of this
Letter. An analogous case is the size of stable domains in
ferromagnets, which cannot be obtained from the Ising Hamiltonian
alone.



\bibitem{Zamolodchikov.cth}
A.B. Zamolodchikov, J.E.T.P. Lett. 43 (1986), 730.



\bibitem{redundant}
This implies that the parameters $g_0, h_0$ couple to the marginal
conformal field $({\bf \nabla} h)^2$ and to a {\em redundant}\ field
that leaves the Gaussian action (\ref{SG}) invariant.



\end{references}
\end{document}